*What do I make of your Latinorum?*
**Sensitivity auditing of mathematical modelling**
*Andrea Saltelli, Ângela Guimarães Pereira,
Jeroen P. van der Sluijs, Silvio Funtowicz*

**Abstract**
In their book 'Uncertainty and Quality of Science for Policy' (1990), Funtowicz and Ravetz argued the need to extend traditional methods and techniques of quality assurance in policy-related science. Since then, these ideas have been operationalized further and applied. Particularly relevant have been the recourse to extended peer review – to be intended as internal, across disciplines, as well as external, between practitioners and stakeholders, and the use of a new approach to qualify quantities: NUSAP (Numeral, Unit, Spread, Assessment, Pedigree). Here we describe how sensitivity analysis, mandated by existing guidelines as a good practice to use in conjunction to mathematical modelling, needs to be transformed and adapted to ensure *quality* in the treatment of *uncertainty* of *science for policy*. We thus provide seven rules to extend the use of sensitivity analysis (or how to apportion uncertainty in model based inference among input factors) in a process of sensitivity auditing of models used in a policy context. Each rule will be illustrated by examples.

# 1. Introduction

In this paper we argue that the quality assessment of mathematical or simulation models that underpin current policy making requires a process which transcends the mere assessment of the model uncertainties and parametric sensitivities (sensitivity analysis) up to include a practice of organized scepticism toward the inference provided by mathematical models (sensitivity auditing). Sensitivity tools need to be adopted which ensure a complete exploration of the space of the input uncertainties, but at the same the boundaries of such a space need to be questioned, if need be also by a process of extended peer review cutting across disciplines as well as across the fence separating practitioners from stakeholders. Sensitivity auditing also needs to cope with non-quantifiable uncertainties, eschewing the hubris of quantification at all cost.

We consider this upgrade of sensitivity analysis as necessary and urgent. On one hand one sees example of instrumental use of mathematical modelling – more to obfuscate that to illustrate, as was the case for the use of Latin by the elites in the classic age (hence the quote in our title). On the other hand as practitioners we are puzzled to see that even when an appraisal of model sensitivities is attempted by modellers, this is often of poor or perfunctory quality (Saltelli and d'Hombres, 2010, Saltelli and Annoni, 2010). In this sense our works feeds into that current of thought which takes issue at the poor quality of existing modelling practices (Taleb, 2007, are Pilkey & Pilkey, 2007, Savage, 2009).

We suggest that through a purposeful organized critical appraisal of model quality that we call "sensitivity auditing", scientific models can better fit the purpose of informing and justifying policy making proposals, in other words make these models more plausible as far their assumptions, outcomes and usage are concerned. In this paper we argue that the focus of such exercise is beyond the model itself but encompasses the entire modelling process. "Sensitivity Auditing" borrows ideas and strategies from sensitivity analysis proper, from post-normal science and from the NUSAP system; it is presented as a set of rules which not only address the pitfalls of the mathematics in the models but also looks into the process of the auditing, in other words not only what to look for but how to look for and who should look for.

So, we start by providing some background of the intellectual context in which we move; we then present what could be seen as a continuous process of "vigilance" that we call sensitivity auditing; we discuss what rules this process should have, including the process by which it could be implemented, and finally we link this endeavour to the central concept of this special issue: plausibility.

## 2. Post Normal Science

In their book 'Uncertainty and Quality of Science for Policy' (1990), Funtowicz and Ravetz developed new conceptual and practical tools for coping with uncertainty and the assurance of quality of quantitative information in policy-related research. Complementing this effort, Funtowicz and Ravetz (199x) also introduced a novel mode of scientific problem-solving appropriate to policy issues where facts are uncertain, values in dispute, stakes high and decisions urgent. They called it Post Normal Science, to relate it to Tomas Kuhn's (1962) book on normal science and to distinguish it from Stephen Toulmin's (1985) postmodern science.

Post Normal Science is a quest for the appropriate management of quality in the presence of irreducible uncertainty (Knight, 1921); it comprises an awareness of the role of values and the acceptance of a plurality of commitments and perspectives. These are expressed through an extended peer community, involving many scientific disciplines, as well as concerned citizens and a plurality of stakeholders in the tasks of problem framing, assessment and quality control.

PNS emerges from the realization that major societal issues involving risk and uncertainty are poorly dealt with by Modern science rigidly organized along disciplinary lines (see also in this respect chapter 9 in Toulmin, 2001), and under the paradigm of "sound science". PNS embraces complexity (including in the set of norms and values) and fosters a new system of scientific governance. The purpose is to enable a plurality of different though legitimate perspectives to be brought to bear on the debate in a reflexive fashion. In this fashion, PNS addresses the so-called type III error (Raifa, 1968; Dunn, 1997), which manifests itself when an issue is framed with the exclusion of one or more relevant and legitimate constituencies.

## 3. NUSAP

Until recently, the field of uncertainty analysis of simulation models mainly evolved around mathematical methods such as error propagation equations and Monte Carlo techniques. These tools address quantitative dimensions of uncertainty. Although these quantitative techniques are essential in any uncertainty analysis, they can only account for what can be quantified and thus provide only a partial insight in what usually is a very complex mass of uncertainties involving technical, methodological, epistemological and societal dimensions. In many cases where models are used to inform policy making, unquantifiable uncertainties may well dominate the quantifiable ones, which implies that these quantitative techniques are of limited value for this particular class of problems.

In the school of Post Normal science, several new multidimensional and reflective approaches have been developed to systematically address unquantifiable dimensions of uncertainty. The most widely known is the NUSAP system for multidimensional uncertainty assessment (Funtowicz and Ravetz, 1990; Van der Sluijs et al., 2005). NUSAP is a tool and notational system for the analysis and diagnosis of uncertainty in science for policy. The basic idea is to qualify quantities - especially those that feed into the policy process - by using the five qualifiers of the NUSAP acronym:

- **Numeral**, the numerical value of the claimed quantity
- **Unit**, its units
- **Spread**, a measure of (statistical or measurement) error
- **Assessment**, an assessment of the reliability of the claim made by experts
- **Pedigree**, which conveys an evaluative account of the production process of the quantity, and indicates different aspects of its underpinning and scientific status.

Pedigree is expressed by means of a set of pedigree criteria to assess these different aspects. Each pedigree criterion has a 5-point scoring scale, the criteria and scores are defined in a pedigree matrix. The attributes used in Pedigree thus represent a multi criteria assessment of the claim and are a function of the type of claim being investigated. If the claim is that some quantity (e.g. northern hemisphere mean temperature in the year 1300) amounts to some numeral (e.g. 14.1°C), the matrix could cover (van der Sluijs, 2005, 2010):

- proxy representation, how close is the primary source of information (e.g. size of tree rings) from which the quantity (e.g. temperature in some past year) was derived to that quantity;
- empirical basis, ranging from 'large sample direct measurement' to 'crude speculation';
- methodological rigor, ranging from 'best available practice' to 'no discernible rigour';
- degree of validation, ranging from 'compared with independent measurements of the same variable' to 'no validation'.

The NUSAP approach is adopted in the Netherlands as part of the Guidance on Uncertainty Assessment and Communication of the Netherlands Environmental Assessment Agency (Van der Sluijs et al., 2008; Petersen et al., 2011).

# 4. Sensitivity Analysis

There is a consensus among practitioners from a plurality of disciplines (Kennedy, 2007; Leamer, 1990; Pilkey and Pilkey-Jarvis, 2007; Saltelli et al., 2008, 2010; Santner et al., 2003; Oakley and O'Hagan, 2004; Saisana et al., 2005) as well as among guidelines devoted to modelling and impact assessment (EC, 2009; EPA, 2009; OMB, 2006) that sensitivity analysis is an indispensable element to judge the quality of inference based on a mathematical models.

Sensitivity analysis' good practices (see a recent review at Saltelli et al., 2005-2012) prescribes that the uncertainty in the inference be quantified by a simultaneous activation of all possible assumptions' uncertainties, followed by an identification of those assumptions chiefly responsible for the uncertainty in the inference. Being numerical experiments, these analyses should be implemented following a statistical design, as one expects for a physical or biological experiment. Assumptions become then factors, whose effect is explored using techniques partly derived from experimental design, a branch of applied statistics.

EPA 2009 describes well what an ideal sensitivity analysis must do:

> *[SA] methods should preferably be able to deal with a model regardless of assumptions about a model's linearity and additivity, consider interaction effects among input uncertainties, [...], and evaluate the effect of an input while all other inputs are allowed to vary as well.*

A class of methods which fulfil EPA's technical requirements is based on decomposing the variance of the inference according to bits which can be attributed to either input factors or combination of factors, the so-called interactions (Saltelli et al., 2008, 2010). This kind of analysis is only successful to the extent that all sources of uncertainties have been identified, which is in most cases impossible to prove (see Rule 7), and that the model is relevant to issue being analyzed. These vast limitations of a technical sensitivity analysis should not justify omitting the analysis of performing it in a perfunctory way. As discusses in Saltelli and Annoni (2010), notwithstanding existing guidelines, most sensitivity analyses seen in the literature tend to display a cavalier attitude with respect to statistical design, model non linearity and model non additivity issues (see Rule 7). As noted in The Flaw of Averages (Savage, 2009) when one scaffold is made by coupling several stairs, one cannot 'shake' one stair at a time to test the safety of the scaffold. A better idea of the stability of the scaffold is obtained by shaking all stairs simultaneously. The fact that books are written to hammer this self-evident point suggests that not all is clear and agreed among practitioners.

# 5. Sensitivity Auditing

Sensitivity auditing aims to extend sensitivity analysis to contexts when mathematical modelling feeds into a policy context. Sensitivity auditing is thus posed to help gauging the quality of scientific information in all cases where models are at play and their outcome feeds into the public discourse, be it in the context of a policy assessment (ex ante or post), or in the general public arenas where policies are contested. Sensitivity auditing starts from the awareness that in an adversarial or media context not only the nature of the evidence, but also the degree of certainty and uncertainty associated to the evidence will be the subject of partisan interests. It encompasses the ideas of post normal science exposed earlier and its associated concept of quality assurance by an *extended peer community*. An *extended peer community* consists not merely of persons with some form or other of institutional accreditation, but rather of all those with a desire to participate in *extended peer review* processes for the resolution of a specific issue. We argue that sensitivity auditing implies in practice the implementation of spaces where relevant social actors are enabled and invited to scrutinise modelling activities (including their policy applications) using their specific knowledge in inclusive and influential ways.

Hence, the set of rules presented in here for sensitivity auditing presupposes that an "extended peer community" is identified and involved in the sensitivity auditing of the mathematical modelling. Useful recipes for sensitivity auditing which are proposed here are:

1. Check against rhetoric use of mathematical modelling;
2. Adopt an 'assumption hunting' attitude;
3. Detect Garbage In Garbage Out (GIGO), in the extended definition of Funtowicz and Ravetz (1990);
4. Find sensitive assumptions before these finds you;
5. Aim for transparency;
6. Do the right sums;
7. Focus the analysis on the key question answered by the model, exploring holistically the entire space of the assumptions.

Before going into the rules in detail, we would like to motivate the need for introducing them with some examples in which modelling used to underpin policy appears dysfunctional.

**1 The financial crisis and the modelling of collateralized debt obligations.**
This is a quite well known story, and concerns the formula of David X. Li and used in the pricing of collateralized debt obligations (the infamous CDO's). The story is popularized in an article of Wired (Salmon, 2009), where it is told how the toxicity of these securities (which packed as many as two thousand individual mortgages into a single obligation) was elegantly overlooked by applying a modelling approach (Gaussian Copula) whereby the probability of joint default of any couple of individual mortgages in the bundle was described by a correlation coefficient estimated on historic data. Unfortunately the 'history' on which this parameter was estimated was a short one, and only relative to a period of housing market up-swing; thus the probability of joint failure of two mortgages was very low in the world of the model. The story changed when the housing bubble exploded, whereby Li's formula lost any predictive power on the world of things. Of course this accident could not have gone overlooked by the 'quants', the mathematicians who are employed in the world of finance. 'Anything that relies on correlation is charlatanism', noted Nassim N. Taleb (cited *ibidem*). The point of the anecdote is that when important stakes are at play the normative stance of all actors – including scientists, must be questioned openly' Yet it would be unfair fingers the quants as the sole modellers with a responsibility for the crisis. As amply debated on the press and in the specialized literature if was the entire macro-economic modelling fabric that was found wanting: "*The standard* *macroeconomic models have failed, by all the most important tests of scientific theory. They did not predict that the financial crisis would happen; and when it did, they understated its effects"* (Stiglitz, 2011). We shall return to the issue in section 6.

**2 Dutch overhead powerlines cause 0.5 cases of childleukemia per year, model says.**

In 2000, the Health Council of the Netherlands reviewed epidemiological state of knowledge on health risks of Extreme Low Frequency Electro Magnetic Field (ELF EMF) and concluded that a 'relatively consistent association between the occurrence of childhood leukaemia and living in the vicinity of overhead power lines' exists. In response, the Ministry asked the Netherlands Institute for Public Health and the Environment (RIVM) to quantify what the risk of overhead powerlines for the Netherlands population would be if one would assume that the association is causal. Making use of estimations on numbers of dwellings in different (magnetic) zones close to overhead power lines, (RIVM) translated the *relative risks* found in international pooled analyses into an annual number of extra cases of childhood leukaemia (Van der Plas et al, 2001; Pruppers, 2003). Their chain of calculations resulted in the claim that overhead power lines add 0.4-0.5 extra cases leukaemia annually in NL (to a total of 110 cases per year). To enable the quantification requested by the Ministry, RIVM had to make a vast amount of assumptions, both prior to and in the model calculation chain. Not all of these were stated explicit in the report. De Jong et al (2012) applied the "assumption hunting" approach to deconstruct RIVM's model calculation. In a first step, 35 assumptions were identified. In an expert workshop that included RIVM experts involved, the list of assumption was reviewed, completed, and ranked according to (estimated) ordinal importance with regard to influence on the outcome of the calculation. The top 5 is listed in table 1. Next, the pedigree of each assumption was assessed. The assumptions with the highest expected impact on the number of extra cases of child leukaemia turned out to be also the ones with the lowest pedigree: many of these assumptions are difficult to underpin and highly value-laden with the state of current knowledge. Moreover, the assumption hunting workshop found that the assumptions which are regarded to be most problematic are prior to the model calculation chain developed by the RIVM: they are hidden in numbers that are taken from other disciplines and fed into the model, such as the *relative risk* factors established in the pooled analysis of epidemiological studies. This finding highlights the key importance of a wide extension of the peer community engaged in model quality control.

| Rank | Assumption |
|---|---|
| 1 | A causal relationship exists between exposure to electromagnetic fields of overhead power lines and the occurrence of childhood leukaemia |
| 2 | Overhead power lines are the main differentiating source of exposure to electromagnetic fields for children |
| 3 | The height of the (prolonged) average of exposure causes the effect |
| 4 | A threshold value exists |
| 5 | The current in the year prior to determining the incidence of childhood leukaemia is representative for the average current during the development of childhood leukaemia |

Table 1: top 5 of assumptions in overhead powerlines health risk study (De Jong et al, 2012)

**3 AIDS**

In their book "Useless Mathematics", Pilkey & Pilkey (2007) describe many instances of dysfunctional policy advice supported by mathematical modelling. We have borrowed this story from their account. Mathematical models can be used to boost causes both bad and good. A troublesome example of good cause modelling is the prediction and monitoring of the spread of HIV AIDS around the world, especially in Africa where the disease is taking its worst toll. UNAIDS takes the responsibility for tracking the disease, which it does in large part through the use of mathematical models. UNAIDS claims 30 M Africans suffer from the disease. In 2003, a South African activist reported in an article Sunday Telegraph that the UN models may have distorted the extent of epidemics in Africa. Quantitative mathematical models are universally used to keep track of and to predict the future courses of diseases but of course models require extensive ground truthing or field checking. In most of southern Africa, record keeping is poor to non-existing, and except for South Africa there is simply no dependable real world information. 250,000 would die in 1999 according to *Epi Model* (an epidemiological model, Chin and Lwanga 1991). But that year 375000 died of all causes. The number of AIDS victims is far too large a proportion, 2/3 of the total deaths.

Another model predicted 143,000 deaths of AIDS. In 2001, the much advanced ASSA 2,000 model predicted that there must have been 92,000 AIDS deaths. There are real difficulties in determining AIDS death rates because the weakened immune system can result in death from a number of causes.

But the experience in South Africa suggests that the AIDS disaster might not be as advanced as previously assumed by the UN, certainly this is a point worth considering because research on other more ravaging diseases in Africa such as Malaria is said to be under-funded because of the anticipated AIDS calamity. (900 000 deaths for Malaria occur every year in sub-Saharan Africa, 70% being children of less than 5 Y). So, where did the numbers come from? There is a genuine lack of records. The models have a poor database.

The possibility that a true global disaster is just around the corner unfortunately provides an unparalleled opportunity for the modelling that checks-up the numbers to draw attention and funding. Failure to make a simple reality check allowed the results to become accepted "facts".

## 6. Rules for Sensitivity Auditing

**Rule 1: Check against rhetoric use of mathematical modelling**

As noted by Hornberger and Spear (1981)

> *[…] most simulation models will be complex, with many parameters, state-variables and non linear relations. Under the best circumstances, such models have many degrees of freedom and, with judicious fiddling, can be made to produce virtually any desired behaviour, often with both plausible structure and parameter values.*

This sober view of modelling was popularized by novelist Douglas Adam in one of his classic novels (1987):

> *Well, Gordon's great insight was to design a program which allowed you to specify in advance what decision you wished it to reach, and only then to give it all the facts. The program's task, […], was to construct a plausible series of logical-sounding steps to connect the premises with the conclusion.*

Adam's irony is cogent. Mathematical modelling is an apt tool to transform evidence based policy in its reverse. The abundance of parameter and assumptions makes the task of mapping the facts to the desired inference trivially easy. This does not apply only to the over-parameterized models addressed by Hornberger and Spear (1981), but also to the relatively parsimonious models used in applied econometrics, as vividly illustrated by Edward E. Leamer (2010).

This use of mathematical modelling (a technique, a language) in a scantly disguised normative (or advocacy) mode can be termed rhetoric, or strategic (Boulanger et al., 2007), like the use of Latin by the elites or the clergy throughout the classic age, in situations where the purpose was to confuse or obfuscate rather than to communicate[1].

There is a vast literature sounding the alarm on instances of corruptions in the use of mathematical models, with the earliest warnings coming by Saunders Mac Lane (1988a,b) in a exchange on letters on the journal SCIENCE on the subject of system analysis as practiced at International Institute for Applied Systems Analysis (IIASA):

> *[…] this type of "systems analysis" consists of the construction of massive imaginary future "scenarios" with elaborate equations for quantitative "models" which combine to provide predictions or projections […] which cannot be verified by checking against objective facts. Instead IIASA studies often proceed by combining in series a number of such unverified models, feeding the output of one such model as input into another equally unverified model.*

The mediatic aspect of the issue is investigated by philosopher Jean Baudrillard (Baudrillard, 1999), according to whom modelling, when used outside '*controlled scientific conditions*' but '*in mass communication, […] assumes the force of reality, abolishing and volatilizing the latter in favour of that neo-reality of a model materialized by the medium itself."*

Along similar lines one of the authors (JvdS, 1998) observed that '*Once environmental numbers are thrown over the disciplinary fence, important caveats tend to be ignored, uncertainties compressed and numbers used at face value*'.

Other recent contributions to the topic are the book of Orrin H. Pilkey and Linda Pilkey-Jarvis (Useless Arithmetic: Why Environmental Scientists Can't Predict the Future), whose title is eloquent enough, and the successful volume of Nassim Nichola Taleb (The Black Swan, 2007), where issue is taken against modelers' attempt to *Platonify* reality, meaning by this the man's attempt to stick to elegant formal structures to describe facts which are too stubborn to be subdued by such simplifications. Taleb's call reminds of Stephen Toulmin's (2001) plea for reasonableness as opposed to overstretched rationality.

---

[1] An illustration is in Alessandro Manzoni's The Betrothed: '*What do you expect me to make of your latinorum? (Che vuol ch'io faccia del suo latinorum)*', is the retort of Renzo, one of the characters in the novel, to the rector of his parish who tries to confuse him with Latin.

Apparently one is never vigilant enough, as --in spite of the warnings cited, many have unfortunately *a posteriori* seen the links between the 2008 credit crunch and the mathematical models disingenuously used to price the financial products at the heart of the crisis. According to Leamer (2010), '*With the ashes of the mathematical models used to rate mortgage-backed securities still smoldering on Wall Street, now is an ideal time to revisit the sensitivity issues*', which incidentally is also the scope of the present work. For Paul Krugman (2010), in a chapter aptly named '*Complexity – going beyond transparency*' notes: '*[…] Part of the agenda of computer models was to maximize the fraction of, say, a lousy sub-prime mortgage that could get an AAA rating, then an AA rating, and so forth, […]*', p. 161, thereby linking '*Perverse incentives*' to '*flawed models*', p.92. Finally Jerry Ravetz (2010), in discussing the ethics of scientists, muses '*Yet we now know that the collective endeavour of these […] very nice entrepreneurial scientists* [the '*quants*', mathematicians employed in finance] *has resulted in the creation of a mountain of toxic fake securities*'.

In summary, Rule 1 prescribes that the prospective sensitivity auditor maintains open eyes and '*organized scepticism*' toward technical and normative hurdles limiting the plausibility of a model based inference.

**Rule 2: Adopt an 'assumption hunting' attitude**.

Models are full of more or less explicit assumptions, which - once made explicit, pose varying challenges to the belief of the beholder. These assumptions may have sedimented into the interstices of a model, or they may have been pondered in the pre-analytic phase of the model and henceforth forgotten by the same users of the model. As an example, in Laes et al., 2011 one notes that in relation to the '*[…] calculation of the external costs of a potential large-scale nuclear accident* […] *[an analysis] resulted in a list of 30 calculation steps and assumptions*'.

Kloprogge et al. (2011) suggest to structure the evaluation of a model based inference into a series of steps covering analysis, revision and communication. The analysis focus on identifying explicit and implicit assumptions in the calculation chain and the potential value-ladennes of key assumptions. The revision includes a sensitivity analysis and a possible diversification of the assumption, while communication aims to make explicit the entire process, inclusive of element of value ladennes and possible alternatives. The degree of value ladennes is estimated via the use of pedigree matrices following the NUSAP methodologies (Funtowicz and Ravetz, 1990). The revision phase combines information from the pedigree analysis and the sensitivity analysis. Those assumptions with a weak pedigree and a strong sensitivity on the inference are those which deserve more scrutiny, revision, and which are at the top of the communication effort.

In the work of Laes et. al. (2011) already cited the communication phase led to a substantial rejection by stakeholders of the model as a relevant tool for policy, as the model's assumptions were judged either implausible or contentious.

An application of these approaches to microbial contamination risk analysis is Boone et al. (2010), while the case of electromagnetic fields is investigated in de Jong et al. (2012).

Just to dispel the impression that assumptions-related scruples are the preserve of a restricted circle of practitioners, one can read on the Financial Times (2011) economist John Kay elaborating on the '*making up*' of the missing data needed to operate models:

> *You assume the future will be like the past, or you extrapolate a trend. Whatever you do, no cell on the spreadsheet may be left unfilled. If necessary, you put a finger in the air. This may lead to extravagant flights of fantasy. To use Britain's Department of Transport scheme for assessing projects, you have to impute values of time in 13 different activities, not just today, but in 2053. [...] What will be average car occupancy rates, differentiated by time of day, in 2035?*

The future being unlike the past (and this being the source of many explicit or implicit assumptions) is an old problem. In the words of Frank Knight (1921):

> *We live in a world of contradiction and paradox, a fact of which perhaps the most fundamental illustration is this: that the existence of a problem of knowledge depends on the future being different from the past, while the possibility of the solution of the problem depends on the future being like the past.*

**Rule 3: Detect garbage in garbage out (GIGO)**

Garbage in garbage out, or GIGO, is the instrumental minimization of uncertainty operated to inflate certainty in the inference, as defined both by econometricians (Edward Leamer, Peter Kennedy) and epistemologists (Silvio Funtowicz and Jerry Ravetz). According to the latters GIGO-science – or pseudo-science, is '*where uncertainties in inputs must be suppressed lest outputs become indeterminate*'. This implies artificially deflating the uncertainty in the assumptions to avoid that the distribution of the inference becomes so flat as to be useless. Saltelli and d'Hombres (2011) use sensitivity auditing to argue that this is the case for the cost benefit analysis proposed by various parties in relation to climate change action or inaction. A very similar standpoint – only turned in an affirmative/positive version, is from Leamer 1990 work, where he states:

> *I have proposed a form of organised sensitivity analysis that I call "global sensitivity analysis" in which a neighborhood of alternative assumptions is selected and the corresponding interval of inferences is identified. Conclusions are judged to be sturdy only if the neighborhood of assumptions is wide enough to be credible and the corresponding interval of inferences is narrow enough to be useful.*

This is after all not a new idea. A related trade-off is in Imre Lakatos (1976, p.57): '*when increasing certainty, you decrease content*', meaning by this that the more one makes a theorem refutation proof, the more the theorem's range of application empties. Leamer's viewpoint is upheld in standard econometrics textbooks; see e.g. Kennedy (2007, see Rule 4).

In a policy context uncertainty can be amplified as well minimized according to convenience. Oreskes and Conway (2010) describe a famous case of uncertainty amplification for tobacco's health effect. These author compare the the narrative of the tobacco companies fighting to deny the health effect of smoking to to those of climate sceptics who – according to these authors – amplified uncertainty about anthropogenic climate change. Naomi Oreskes is well known to modellers for having extensively written against the concept of model validation or verification. According to Oreskes, models can be evaluated, or corroborated, but never be proven true (1994).

In a later work Oreskes (2000) articulated her critique by noting that

> *models are complex amalgam of theoretical and phenomenological laws (and the governing equations and algorithms that represent them), empirical input parameters, and a model conceptualization. When a model generates a prediction, of what precisely is the prediction a test? The laws? The input data? The conceptualization? Any part (or several parts) of the model might be in error, and there is no simple way to determine which one it is.*

Oreskes's point is linked to the parallel often made between a logical proposition – a theory-based statement - and a model prediction. Although models share the scientific flavour of postulated laws or theories they are not laws in that the making of a model is substantially more fraught with assumptions than crisp theories or agile laws ordinarily are. She notes '[…] *to be of value in theory testing, the predictions involved must be capable of refuting the theory that generated them.*' What when the 'theory' is not a law but a mathematical model? '*This is where predictions [...] become particularly sticky.*' The crux of the matter is that model based inferences are very delicate artefacts.

Another interesting story about uncertainty manipulation is that told by David Michaels – a former EPA employee, on the battles between industry and regulators over the US data quality act and the standard for exposure to beryllium. This is where industry fought hard to amplify uncertainty, according to the author, as to prevent regulators from imposing more stringent standards. The same debate in the US surrounded the Proposed Risk Assessment Bulletin (January 9, 2006, see http://www.whitehouse.gov/omb/inforeg/, which was received by some as an attempt '*to bog the* [regulatory] *process down, in the name of transparency'* (Robert Shull cited in Macilwain, 2006). In the same article one reads ' *[...] the proposed bulletin resembles several earlier efforts, including rules on 'information quality' and requirements for cost–benefit analyses, that make use of the OMB's* [Office for Management and Budget] *extensive powers to weaken all forms of regulation.*'

An important consequence of Rule 3 is that one should particularly severe against spurious accuracy, e.g. when a result is given with a number of digits exceeding (at time ludicrously) a plausible estimate of the associated uncertainty.

**Rule 4: Find sensitive assumptions before these finds you**

One of the ten commandments of applied econometrics according to Peter Kennedy popular Econometrics textbook on Applied Econometrics is: '*Thou shall confess in the presence of sensitivity. Corollary: Thou shall anticipate criticism.*' This wisdom of this principle is evident, in that when an unwanted or unexpected model sensitivity is exposed by a third party, it becomes arduous for the proponent modellers to reinstate a just-falsified inference. Thus sensitivity analysis, or better sensitivity auditing, can be used to anticipate a critique. This is the application to modelling of Robert K. Merton '*organized scepticism*'. According to Merton *Communalism, Universalism, Disinterestedness and Organized Skepticism* are the operating principles (norms) of the scientific method. Also for Pilkey and Pilkey-Jarvis (2007) '*Scientific mathematical modelling should involve constant efforts to falsify the model*'.

**Rule 5: Aim to for transparency**.

The discussion of Rule 3 above about the data quality act and the risk assessment bulletin has shown how the issue the issue of transparency can be the subject of dispute. While transparency can in general be seen as an element of quality, it can at times be denounced as pretext to '*bog the process down*'. While keeping this caveat in mind, we shall mostly embrace transparency as useful in the context of mathematical modelling when this has to feed into the policy process.

According to the OMB (2002) models should be made available to a third party so that it can '*use the same data, computer model or statistical methods to replicate the analytic results reported in the original study. […] The more important benefit of transparency is that the public will be able to assess how much an agency's analytic result hinges on the specific analytic choices made by the agency.*'

The OMB suggestion hence is that reproducibility is an necessary condition to transparency. Our suggestion is that transparency is in turn useful to defend the legitimacy and epistemic authority of the institutions making use of mathematical modelling in the context of a policy assessment.

Often the same model used within an organization to paddle through the analysis, the workhorse pulling the cart of the daily 'what if', '*caeteris paribus*' work, might be used in an adversarial context, simply because it is expected that external stakeholder will accept the house's wisdom and its model. This may well be the case, but one would be wise not to bank on it. The problem is that in real life *caeteris* are never *paribus*.

In the words of Joseph Stiglitz: "*Models by their nature are like blinders. In leaving out certain things, they focus our attention on other things. They provide a frame through which we see the world.*" (2011)

A considerable amount of work is needed to transform the workhorse (*the frame through which we see the world*) into something more agile and proportionate that can stand in court. Real life examples show that model use may even become counter productive in a policy debate when this type of simplification is not operated. In the context of climate this point is made by Pilkey and Pilkey-Jarvis (2007, chapter 4), where it is argued that the climate-sceptics' work would be harder if *[…] the global change modeling community would firmly and publicly recognize that its efforts to truly quantify the future are an academic exercise and that existing field data on atmospheric temperatures, melting glaciers, […] and other evidence should be relied on to a much greater degree to convince politicians that we have a problem. Let the models point to a trend and answer 'what-if' questions. A serious societal debate about 'solutions' can never occur as long as modellers hold out the probability, just around the corner, of accurate projections of future climates and seal-level position.*

Five years after the publication of this book we see comforting signs that the point has been driven home; it is now admitted that the more one understands climate, the more model predictions may become uncertain (Maslin and Austin, 2012), and more and more means and standard deviations (e.g. of temperature) populate the discourse on climate (Hansen, et al., 2012). Still policy makers associate a 50% certainty to meeting a 2 degree centigrade temperature increase, a climate policy target, with a greenhouse gas concentrations at 450 ppm CO2-equivalent (Meinshausen et al., 2005[2]). Given that these three numbers (0.5, 450, 2) are model-generated some more circumspection would befit the prospective sensitivity auditor.

Finally one had to admit that at present there is in general little scientific incentive to reproduce a model. This has been observed in scientific papers (recently in Nature [reference]): nobody tries to reproduce the results because it takes resources that are never justified. The only arena in which it is done is in an already contested and controversial issue and where the stakes are no longer in the context of 'normal' science but in the context of use.

**Rule 6: Do the right sums**

In 'Return to Reason' Stephen Toulmin vividly recalls the dangers of precision: '*Doing the sum right*' is a far lesser challenge than '*Doing the right sums*'.
In modelling as in life framing error, or Type 3 errors, are the most dangerous. When performing an uncertainty and sensitivity analysis one falls easily into that which N. Taleb (2007) calls '*The delusion of uncertainty*' and which is known in Dutch as '*Lampposting*', whereby '*The uncertainties which are more carefully scrutinised are usually those which are the least relevant*, (van der Sluijs, reference missing). Lampposting refers to the joke of the drunkard looking for his lost keys not in his house's garden, where he lost them, but in the street under a lamp as '*there is more light*'.

A type three error is illustrated in the work of Pilkey and Pilkey-Jarvis already cited, The examples concerns the Yucca Mountain repository for radioactive waste. A model named TSPA (for total system performance assessment) has been used for the safety analysis computations. TSPA is Composed of 286 sub-models. A key assumption in TSPA is the range of permitted values for the permeability of the geological formation. A low permeability is key to ensure that water will take a long time to percolate from surface to disposal. For the Yucca mountain test disposal site a range of 0.02 to 1 millimetre per year was used for the percolation rate. The confidence of the stakeholders in TSPA was not helped when evidence was produced which led to an upward revision of 4 orders of magnitude of this parameter (order of three metres per year). The evidence in question was the presence at the repository level of an isotope of chlorine $^{36}Cl$ associated to atomic bombs detonations in the atmosphere. According to the authors the error was due to the modelling of the granite formation as a homogeneous medium, while a fissures and faults model of the same formation would have been more realistic.

---

[2] http://www.pik-potsdam.de/~mmalte/simcap/publications/meinshausenm_risk_of_overshooting_final_webversion.pdf

Sensitivity analysis is no immune to type three errors, and neither is sensitivity auditing (Pilkey and Pilkey-Jarvis, 2007):

> *'It is important […] to recognize that the sensitivity of the parameter in the equation is what is being determined, not the sensitivity of the parameter in nature. […] If the model is wrong or if it is a poor representation of reality, determining the sensitivity of an individual parameter in the model is a meaningless pursuit'*

Type three error which are most common are likely those associated to neglecting part of the views and expectations surrounding the issue at stake.

**Rule 7: Focus the analysis**

Sensitivity is often omitted in modelling studies, or it is executed in a perfunctory fashion. According to Leamer (1010) '*One reason these methods* [global sensitivity analysis] *are rarely used is their honesty seems destructive*'. Most sensitivity analysis seen in the highest ranking journals such *SCIENCE* and *Nature* are perfunctory. The may sound a surprising claim. Still the analyses of sensitivity reviewed in Saltelli and Annoni (2010) were run without a statistical design, moving just one input factor at a time. Beside neglecting the existence if interactions among factors, this approach bumps against the curse of dimensionality, as at already moderate dimensionality moving one factor at a time explores only a tiny fraction of the space of the input.

In general a sensitivity analysis performed without a statistical design and without an estimate of the error is poor. In the context of modelling studies used in support to policy sensitivity analysis should also be parsimonious and possible cogent, i.e. it should focus on a single target variable, being this the relevant inference that the modelling study is trying to underpin. The analysis should be one and not many, covering to the entire evidential chain, as opposed to covering one sub-model at a time. Again this is needed to ensure that all interactions among factors in different compartments are being captured. Following this view, rather than diluting the sensitivity analysis showing its results for different scenarios, the scenario should be one of the variables activated in the frame of a single analysis.

An illustration of Rule 7 to the sensitivity analysis performed in the context of the Stern Review is in Saltelli and d'Hombres (2012). In this paper one criticises both the authors of the Stern review and their opponents for lacking a rigorous design-based sensitivity analysis and for using highly uncertain numbers (e.g. discount rates) at face value extending the scope of a cost benefit analysis to cover two centuries from present time, thus appearing to use mathematical models expediently, for the sake of projecting pre-established normative stances.

As argued earlier, the process by which sensitivity auditing is carried out should be consistent with the ideas of post-normal science embraced in here. So, a participatory process where *relevant* members of the extended community of peers are first identified and subsequently involved in this process needs to be organised. The identification of this community can be done in many ways, the most obvious being through institutional analysis and stakeholder analysis. The spaces where such scrutiny occurs can be a myriad according to the communities involved. Such spaces, safe and authorised (Guimarães Pereira et al. 2010) take the vest of "focus groups", "juries", consensus workshops, or merely in-depth interviews with specific individuals. It is obvious that the less specialised is the community involved the better the unfolding of the modelling "black boxes" has to be prepared. We argue that "progressive disclosure of information" (Guimarães Pereira et al. 2006) is a key principle of design of communication in participatory settings due to the support it provides for mixed *expertises*. Information is supplied in layers of increasing specialisation, depending on actors' interests and necessary specialised information. So, in other words on a process of auditing such as the one suggested, the modelling "black box" needs to be progressively and intelligibly open to those who participate in the exercise. The community involved, the space and the fairness with which the object of scrutiny is looked at gives the legitimacy to the whole exercise.

## 7. Conclusions

Throughout this paper we have discussed of mathematical modelling in general, without distinguishing between data driven and principle driven models, between micro and macro, between natural sciences and social sciences styles of modelling. It is clear that the arguments developed in this paper are general to forms of evidence that demand statistical, mathematical or otherwise disciplinary elaboration. We wold hence prescribe similar recipes when the model is in fact a statistical indicator, whose construction customarily demands several modelling steps (Boulanger et al, 2007, Paruolo et al., 2012).

Since no recipe is universal, many words of caution should be added to our list of prescriptions, however sensible these may appear to a benevolent reader. In a value-laden context the correctness of the prescriptions may well be a casualty in the power game, as the first casualty of a battle is the battle plan.

Indeed the quality of the rules can only be judged in relation to their fitness for an assigned purpose in a specified case, and there is no guarantee that major blunders will be avoided by a diligent application. The rules are a minimum, due-diligence requirement for the use of model based inference in a policy discourse, and there seem to be little justification, given the stakes involved in any policy, in omitting these simple well meaning steps. At the very least, sensitivity auditing will ensure that the recipients of the analysis are fully aware of the conditionality of the predictions, and a notch more sceptical of model-based evidence when this is presented on the basis of an authority principle.

It is possible that the practitioners' community is becoming more sympathetic with the 'uncertainty exploration' concerns raised in the present work, even in hotly debated areas such as climate. "*Many of those of us who spend our working hours, and other hours, thinking about uncertainty, strongly believe the climate modelling community must not put resolution and processes (to improve the simulator) above generating multiple predictions (to improve our estimates of how wrong the simulator is).*" Our optimism rest nevertheless tempered by the fact that the quote just offered originates from a blog (allmodelsarewrong.com, Edwards 2012) whose title puts it squarely in the field of sympathizers to sensitivity auditing.

PNS likewise promotes awareness of the '*danger that public policy could itself become the captive of a scientific-technological elite*' (Eisenhower, 1961).

Link to plausibility
Why is this relevant for plausibility?
Could the result of this auditing be express with ideas of plausibility?
- the applause of many; so a shared agreement for example on rhetoric, assumptions; what numbers to use, what sums to carry out, etc.
- Plausibility after all maps onto fitness for purpose.